\documentclass[12pt]{iopart}

\usepackage{iopams}  
\usepackage{graphicx}
\begin{document}

\title[Glauber dynamics of a bond-diluted Ising model on a Bethe lattice]
{Multi-step relaxations in Glauber dynamics of a bond-diluted Ising model on a Bethe lattice}

\author{Hiroki Ohta}

\address{Department of Pure and Applied Sciences,
University of Tokyo, 3-8-1 Komaba Meguro-ku, Tokyo 153-8902, Japan}
\ead{hiroki@jiro.c.u-tokyo.ac.jp}
\begin{abstract}
Glauber dynamics of a bond-diluted Ising model on a Bethe lattice 
(a random graph with fixed connectivity) is investigated by 
an approximate theory which provides exact results 
for equilibrium properties.
The time-dependent solutions of the dynamical system derived by this method 
are in good agreement with the results obtained by Monte Carlo simulations in
almost all situations.
Furthermore, the derived dynamical system exhibits
a remarkable phenomenon that 
the magnetization shows multi-step relaxations at intermediate time scales 
in a low-temperature part of the Griffiths phase 
without bond percolation clusters.
\end{abstract}

\maketitle

\newcommand{\bra}{\left\langle}
\newcommand{\ket}{\right\rangle}
\newcommand{\pder}[2]{\dfrac{\partial #1}{\partial  #2}}
\newcommand{\pdert}[2]{\dfrac{\partial^2 #1}{\partial  #2^2}}
\newcommand{\der}[2]{\dfrac{d #1}{d  #2}}
\newcommand{\bv}[1]{{\boldsymbol #1}}
\newcommand{\p}{\partial_t}
\newcommand{\ovl}[1]{\overline #1}

\newcommand{\hc}{h_{\rm c}}
\newcommand{\rhov}{\rho_{\rm v}}
\newcommand{\nug}{\nu_{\rm g}}
\newcommand{\num}{\nu_{\rm m}}
\newcommand{\Nc}{N_{\rm c}}
\newcommand{\Dt}{\Delta t }
\newcommand{\thetas}{\theta^{\rm s}}
\newcommand{\Domh}{{\rm Dom}_h}
\newcommand{\rhoq}[1]{\rho_{#1}^{\rm q}}

\newcommand{\du}{d_{\rm u}}
\newcommand{\Rcsp}{R_{\rm c}^{\rm sp}}
\newcommand{\Rceq}{R_{\rm c}^{\rm eq}}
\newcommand{\rhos}{{\boldsymbol  \rho}_{\rm c}}
\newcommand{\ep}{\epsilon}
\section{Introduction}
%%% impurity in statistical physics: Griffiths-McCoy singularity

Clarifying the role of impurities in many-body systems is a central issue 
in statistical physics 
because statistical quantities of a system with a kind of impurities 
are often different from those of a system without the impurities qualitatively.
Griffiths-McCoy singularities, corresponding to the essential singularity 
appearing in the density of Lee-Yang zeros of a partition function, are 
such typical phenomena \cite{Griffiths,McCoy,Vojta}.

%%% previous studies and unsolved problem related to GM singularity
Thus far, the equilibrium properties of Griffiths-McCoy singularities
have been extensively investigated \cite{Vojta,Hukushima}.
In order to simplify the problem on finite-dimensional systems, 
models on Bethe lattices have been studied
with special theoretical techniques, 
which accelerate the understanding of the equilibrium properties 
of the singularities \cite{Harris,Bray1,Barata,Laumann}.
However, in contrast to the considerable amount of knowledge available on 
equilibrium properties, the knowledge available on dynamical properties 
associated with Griffiths-McCoy singularities is scarce except for 
a few previous studies which have clarified anomalous dynamical behaviours 
in the long-time limit \cite{Bray2,Maes1,Maes2}. 
Thus, the knowledge of dynamical behaviours at intermediate time scales 
has not been established although such a challenge for a model on a Bethe lattice exists \cite{Coolen1}. However, even if a Bethe lattice is adopted as a simple 
situation, 
the understanding of dynamical behaviours at intermediate time scales is still limited. 
Probably, one of the reasons for this limited understanding is 
the lack of theoretical treatments for such dynamical problems.

%%% focused problem and obtained results in this paper
In this study, in order to solve such a problem, 
we attempt to develop an approximate theory
for analysing the dynamics of a bond-diluted Ising model on a regular random graph 
with fixed connectivity (a Bethe lattice). 
We find that the final states of the derived dynamical system are exact, 
and the time-dependent solutions of the dynamical system 
are in good agreement with the results obtained by Monte Carlo (MC) simulations
at almost all situations, 
although there are slight discrepancies between the two results 
below the Griffiths-paramagnet transition temperature with 
bond percolation clusters. 
From the results of this method, we predict that 
magnetization shows multi-step relaxations at intermediate time scales 
in a low-temperature part of the Griffiths phase 
without bond percolation clusters.

%%% outline of this paper
This paper is organized as follows. In section 2, we define 
a bond-diluted Ising model precisely.
In section 3, we derive a dynamical system from the model approximately. 
In section 4, we investigate the concrete behaviours of 
the derived dynamical system and compare them with MC simulations. 
In the concluding remarks, we discuss the validity of the approximate method.
In the Appendix, we present the equilibrium properties of the bond-diluted Ising model on the Bethe lattice.

\section{Model}\label{model}
Let us consider a regular random graph, which consists of $N \in\mathbb{N}$ sites, and each site connects to $c \in \mathbb{N}$ sites chosen randomly where $\mathbb{N}$ is the set of natural numbers.
Then, ${\rm G}(c,N)$ is defined as a set of such regular random graphs. 
For the spin variable $\sigma_i\in \{-1,1\}$ defined on each site $i\in \{1,\cdots,N\}$ in a random graph $\mathcal{G}\in {\rm G}(c,N)$, 
we consider a bond-diluted Ising model, whose Hamiltonian is given as 
\begin{eqnarray}
H^{\bv{J}}(\bv{\sigma}) = -\frac{1}{2}\sum_i\sum_{j\in B_i}J_{ij}\sigma_i\sigma_j,
\end{eqnarray} where $B_i$ is defined as a set of sites connected to site $i$,
and we collectively express $\bv{\sigma}\equiv(\sigma_i)_{i=1}^N$ and $\bv{J}\equiv(J_{ij})_{i,j=1}^N$. 
The configuration of bonds $J_{ij}\ (=J_{ji})$ is given by the probability
\begin{eqnarray}
D(J_{ij})= p \delta(J_{ij}-1)+ (1- p)\delta(J_{ij}), 
\end{eqnarray} for $j\in B_i$, otherwise $J_{ij}=0$. 
The average of a quantity $X$ over diluted bonds is expressed as $\overline{X}$.
Next, we define the dynamics of the bond-diluted Ising model as follows.
Let $T^{\bv{J}}(F_i\bv{\sigma}\to\bv{\sigma})$ be 
the transition rate from $F_i\bv{\sigma}$ to $\bv{\sigma}$ 
which satisfies the detailed balance condition, 
where $F_i$ is the spin flip operator at site $i$ such that
$F_i\bv{\sigma}=(\sigma_1,\cdots,-\sigma_i,\cdots,\sigma_N)$. 
The master equation for the probability $P^{\bv{J}}(\bv{\sigma},t)$, 
that the spin configuration is $\bv{\sigma}$ at time $t$ with a realization of 
$\bv{J}$, is as follows:
\begin{eqnarray}
\partial_t P^{\bv{J}}(\bv{\sigma},t)=\sum_{i=1}^N[T^{\bv{J}}(F_i\bv{\sigma}\to\bv{\sigma}) 
P^{\bv{J}}(F_i\bv{\sigma},t)-T^{\bv{J}}(\bv{\sigma}\to F_i\bv{\sigma})P^{\bv{J}}(\bv{\sigma},t)]. \label{micmas}
\end{eqnarray}
In this paper, we consider the case
\begin{eqnarray}
T^{\bv{J}}(F_i\bv{\sigma}\to\bv{\sigma})=
\frac{1}{2}[1-\tanh((H^{\bv{J}}(F_i\bv{\sigma})-H^{\bv{J}}(\bv{\sigma}))/2T)].
\end{eqnarray}

\begin{figure}
\centering
\includegraphics[width=7cm,clip]{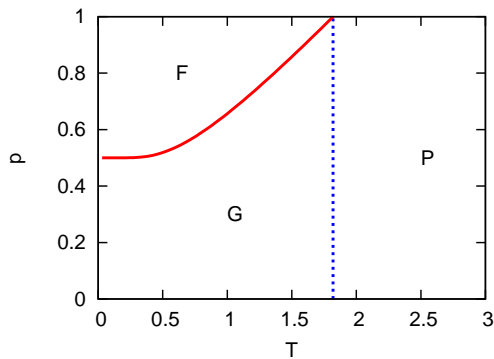}
\caption{Phase diagram of the system with $c=3$ where 
F, P and G indicate the ferromagnetic phase, paramagnetic phase and 
Griffiths phase, respectively.}
\label{static}
\end{figure}

Here we briefly review the equilibrium properties of Griffith-McCoy singularities.
Let $Z(\theta)$ be a partition function of a pure imaginary external field $\theta$.
The term of Griffiths-McCoy singularities is defined as 
the fact that in the thermodynamic limit,
the density of zeros of $Z(\theta)$ in the imaginary axis has the essential singularity 
such as $\exp(-A(p,T)/|\theta|)$ in the limit $|\theta|\to 0$, 
where $A(p,T)$ is a constant depending on the values of parameters $(p,T)$. 
It should be noted 
that with the values of the parameters at which such singularities occur, 
the magnetization in equilibrium with $\theta=0$ is still zero. 
The Griffiths phase where such singularities are observed is shown in figure \ref{static}. 
The phase boundary in terms of the ferromagnetic phase is obtained from the discussion presented in Appendix A. 
It should be noted that the phase boundary between the Griffiths phase and the paramagnetic phase 
is not obtained by the discussion presented in Appendix A.
More details about Griffiths-McCoy singularities are discussed in \cite{Laumann}.
Naively speaking, Griffiths-McCoy singularities originate from 
rare large ordered regions. The main focus of this study is to understand
the anomalous dynamical behaviours associated with the singularities. 
In the next section, 
we construct an approximate theory to understand such behaviours.

\section{Derivation of an effective dynamical system}
In principle, master equation (\ref{micmas}) provides complete information 
about the system. However, it is difficult to extract useful information 
from this equation
 because the number of states of the system is $2^N$, 
which is quite large when $N$ is large. 
Furthermore, the analysis of the system becomes more complicated 
because of the presence of diluted bonds as a quenched disorder.
Here, let us remind that a useful theoretical method has been constructed to 
understand equilibrium properties of some systems on Bethe lattices without directly considering such large number of variables \cite{Parisi}.
On the basis of the effectiveness of this method, first, 
we attempt to describe the system with a set of 
a finite number $N_{\rm eff} \in O(N^0)$ variables 
in order to obtain more useful information on the dynamical behaviours 
of the system, 
However, in general, it is not straightforward to extend the exact analysis 
from equilibrium properties to dynamical properties.
In fact, although some works have been succeeded 
in deriving the exact evolution equations for a reduced number of variables 
\cite{Majumdar1,Majumdar2,Iwata,Mimura,Neri,Ohta1}, 
each method has each special aspect, 
by which it is difficult to obtain 
useful information about the dynamical properties of the system 
in the case of the present problem.
As another way to obtain them, 
let us focus on a remarkable previous study 
\cite{Semerjian}. The study has reported that 
an approximate description with a finite number of variables, which is exact 
for equilibrium states, can be obtained in an Ising ferromagnet. 
Here, on the basis of the results of this study, 
we attempt to develop an approximate theory
for dealing with the quenched disorders of the system 
on Bethe lattices. 

First, as a preliminary step, we call site $j\in B_i$ with the bond $J_{ij}=1$ 'bonded site of site $i$', and we call site $j\in B_i$ with the bond  $J_{ij}=0$ 'unbonded site of site $i$'. The number $l_i \in \{0,\cdots,c\}$ 
of bonded sites of site $i$ is determined by 
\begin{equation}
l_i=\sum_{j\in B_i}J_{ij}.
\end{equation} 
Let $u_i\in\{0,\cdots,l_i\}$ denote the number of downward spins 
at the bonded site of site $i$. 
Then, site $i$ is characterized by a set of variables 
$(\sigma_i,u_i,l_i)$. 
Using these variables, 
we can express $T^{\bv{J}}(\bv{\sigma}\to F_i\bv{\sigma})$ 
as $r_{\sigma_i}(l_i,u_i)$, 
where $r_{\sigma}(l,u)\equiv(1-\tanh(\sigma(l-2u)/T))/2$.
It should be noted that $\sigma_i$ and $u_i$ depend on time $t$, 
whereas $l_i$ does not depend on time $t$.
Next, with a time $t$ and a realization of diluted bonds $\bv{J}$, 
let $P_i^{\bv{J}}(\sigma,u;t)$ be the probability that $(\sigma_i,u_i)$ 
takes $(\sigma,u)$, and let $P_{ij}^{\bv{J}}((\sigma,u),(\sigma',u');t)$ 
be the joint probability that $(\sigma_i,u_i)$ and $(\sigma_j,u_j)$ take 
$(\sigma,u)$ and $(\sigma',u')$, respectively. 
Using these probabilities, we also define the conditional probability 
$P_{ij}^{\bv{J}}(\sigma,u|\sigma',u';t)\equiv P_{ij}^{\bv{J}}((\sigma,u),(\sigma',u');t)
/P_j^{\bv{J}}(\sigma',u';t)$.
In the following analysis, we fix a graph $\mathcal{G}\in{\rm G}(c,N)$ for 
sufficiently large $N$ without considering the ensemble for ${\rm G}(c,N)$.
That is, the following analysis can be applicable to almost all graph 
$\mathcal{G}\in {\rm G}(c,N)$ in the thermodynamic limit.

With these notations, we start with the following exact expression:
\begin{eqnarray}
\partial_t P_i^{\bv{J}}(\sigma,u;t) &=- r_{\sigma}(u,l_i)P_i^{\bv{J}}(\sigma,u;t) + r_{-\sigma}(u,l_i)P_i^{\bv{J}}(-\sigma,u;t) \nonumber\\
&+\sum_{j\in B_i}\sum_{\sigma'}\sum_{u_j=0}^{l_j}r_{\sigma'}(u_j,l_j)P_j^{\bv{J}}(\sigma',u_j;t)\delta(J_{ij}-1)\nonumber\\
&[P^{\bv{J}}_{ij}(\sigma,u+\sigma'|\sigma',u_j;t)-P_{ij}^{\bv{J}}(\sigma,u|\sigma',u_j;t)], \label{mas1}
\end{eqnarray} where we used the property that 
the order of the length of loops in the random graph ${\mathcal G}$ 
is $O(\log N)$ longer than three, 
and we modified the definition $P^{\bv{J}}(\sigma,u|\sigma',c+1;t)=P^{\bv{J}}(\sigma,u|\sigma',-1;t)\equiv 0$ for convenience.
We carry out an average over diluted bonds by multiplying both the sides of equation (\ref{mas1}) 
with $\delta(\sum_{j\in B_i}J_{ij}-l)$. Here, we define 
\begin{eqnarray}
P_i(\sigma,l,u;t)\equiv\overline{\delta(\sum_{j\in B_i}J_{ij}-l)P_i^{\bv{J}}(\sigma,u;t)},\\
P_{ij}(\sigma,l,u|\sigma',l',u';t)\equiv\nonumber\\
\overline{\delta(J_{ij}-1)\delta(\sum_{k\in B_i}J_{ik}-l)
\delta(\sum_{k'\in B_j}J_{jk'}-l')P_{ij}^{\bv{J}}(\sigma,u|\sigma',u';t)},\\
P_j^i(\sigma,l,u;t)\equiv\overline{\delta(J_{ij}-1)\delta(\sum_{j\in B_i}J_{ij}-l)P_i^{\bv{J}}(\sigma,u;t)}.
\end{eqnarray}
Then, we have
\begin{eqnarray}
\partial_t P_i(\sigma,l,u;t) &= - r_{\sigma}(l,u)P_i(\sigma,l,u;t) + r_{-\sigma}(l,u)P_i(-\sigma,l,u;t) \nonumber\\
&+ \sum_{j\in B_i}\sum_{\sigma'}\sum_{u'\le l'}r_{\sigma'}(l',u')
P_j^i(\sigma',l',u';t)\nonumber\\
&[P_{ij}(\sigma,l,u+\sigma'|\sigma',l',u';t)-P_{ij}(\sigma,l,u|\sigma',l',u';t)], \label{mas2}
\end{eqnarray} where $\sum_{u'\le l'}\equiv\sum_{l'=0}^c\sum_{u'=0}^{l'}$.
Key identities for the derivation of (\ref{mas2}) from equation (\ref{mas1}) are
\begin{eqnarray}
\overline{A_i}=\sum_{l'=0}^c\overline{\delta(\sum_{j\in B_i}J_{ij}-l')A_i}, \\
\overline{\delta(a)X(a)Y(a)/Z(a)}=\overline{\delta(a)X(a)}
\frac{\overline{\delta(a)Y(a)}}{\overline{\delta(a)Z(a)}}. \label{rel}
\end{eqnarray} 

Hereafter, we assume that the values of $P_j^i(\sigma',l',u';t)$ and 
$P_{ij}(\sigma,l,u'|\sigma',l',u';t)$ do not depend on 
the two chosen bonded sites $i,j$ in the thermodynamic limit.
This assumption may be plausible because 
at least, in this case, inhomogeneous properties of the system 
originated from the effects of loops of the random graph $\mathcal{G}$ 
may be negligible in the thermodynamic limit.
This assumption corresponds to that $P_{ij}(\sigma,l,u'|\sigma',l',u';t)$ is
the same as the conditional probability $P_{\rm B}(\sigma,l,u|\sigma',l',u';t)$ 
that if a site characterized by $(\sigma,l,u)$ is randomly chosen, 
another site characterized by $(\sigma,l,u)$ is randomly chosen 
in the bonded sites of the first chosen site, 
and also that $P_j^i(\sigma',l',u';t)$ is equal to $(l'/c)P_j(\sigma',l',u';t)$.
Here, we define $P_{\rm B}^{\sigma}(l,u|\sigma',l',u';t)$ as follows:
\begin{eqnarray}
P_{{\rm B}}^{\sigma}(l,u|\sigma',l',u';t)\equiv\frac{P_B(\sigma,l,u|\sigma',l',u';t)}{P_{\rm B}(\sigma|\sigma',l',u';t)},
\end{eqnarray}
where 
$P_{\rm B}(\sigma|\sigma',l',u';t)$ is defined by the similar way 
to that of $P_{\rm B}(\sigma,l,u'|\sigma',l',u';t)$. 

Then, we can obtain the following relation:
\begin{eqnarray}
P_{\rm B}(\sigma|\sigma',l',u';t)&=
\frac{C_\sigma(l',u')}{l'}, \label{num}\\
C_\sigma(l',u')&=\left\{
\begin{array}{ll}
l'-u' & (\sigma=1)\\
u' & (\sigma=-1).
\end{array}\right.
\end{eqnarray} 
In the above procedures with a trivial relation $(\sum_{j\in B_i})=c$, dynamical system (\ref{mas2}) is simplified into
\begin{eqnarray}
\partial_t \rho_{\sigma,l,u}(t) &=- r_\sigma(l,u)\rho_{\sigma,l,u}(t) + r_{-\sigma}(l,u)\rho_{-\sigma,l,u}(t) \nonumber\\
&+\sum_{\sigma'}\sum_{u'\le l'}r_{\sigma'}(l',u')\rho_{\sigma',l',u'}(t)C_\sigma(l',u')\nonumber\\
&[P_{\rm B}^{\sigma}(l,u+\sigma'|\sigma',l',u';t)-P_{\rm B}^{\sigma}(l,u|\sigma',l',u';t)], \label{dyn1}
\end{eqnarray} where $\rho_{\sigma,u,l}(t)\equiv\sum_{i=1}^NP_i(\sigma,u,l;t)/N$. 
It may be plausible that 
$\rho_{\sigma,u,l}(t)$ is identical to
$\sum_{i=1}^N\delta(\sigma-\sigma_i(t))\delta(u-u_i(t))\delta(l-l_i)/N$ 
in the thermodynamic limit $N\to\infty$. 
At the present level, equation (\ref{dyn1}) is not closed in terms of 
$\bv{\rho}\equiv (\rho_{\sigma,l,u})$. 

Next, in order to obtain a closed equation based on (\ref{dyn1}), 
we carry out the following approximation in equation (\ref{dyn1}):
\begin{eqnarray}
P_{{\rm B}}^{\sigma}(l,u|\sigma',l',u';t)= P_{{\rm B}}^{\sigma}(l,u|\sigma';t).\label{rep}
\end{eqnarray}
By considering the number of connected sites on which the spin variable 
is $\sigma'$, we can express $P_{{\rm B},\sigma}(l,u|\sigma';t)$ using $\bv{\rho}$ as follows:
\begin{eqnarray}
P_{{\rm B}}^{\sigma}(l,u|\sigma';t) =
\left\{
\begin{array}{ll}
 (l-u)\rho_{\sigma,l,u}(t)/\sum_{u'\le l'}\rho_{\sigma,l',u'}(t)& (\sigma'= +1)\\
u\rho_{\sigma,l,u}(t)/\sum_{u'\le l'}u'\rho_{\sigma,l',u'}(t)& (\sigma'=-1).\label{rel4}
\end{array}\right.
\end{eqnarray} 
From (\ref{dyn1}) with approximation (\ref{rep}), 
we obtain a closed evolution equation in terms of $\bv{\rho}$ as 
follows:
\begin{eqnarray}
\partial_t \bv{\rho}=\bv{G}(\bv{\rho}), \label{dyn}
\end{eqnarray} where $\bv{G}=(G_{\sigma,l,u})$, and $G_{\sigma,l,u}$ is a function expressed 
by the right-hand side of equation (\ref{dyn1}) with approximation (\ref{rep}) 
and relation (\ref{rel4}). At the present level, 
the number of system variables is reduced to $N_{\rm eff}=(c+1)(c+2)/2$ from $2^N$.

In order to consider the validity of approximation (\ref{rep}), first, we focus on
the stationary solution $\bv{\rho}_{\rm eq}$ satisfying $\bv{G}(\bv{\rho}_{\rm eq})=0$ 
and define the time $t_{\rm eq}\equiv \min t_{\rm st}$ such that 
$\bv{\rho}(t_{\rm st})=\bv{\rho}_{\rm eq}$.
By rewriting 
\begin{eqnarray}
P_{{\rm B}}^{\sigma}(l,u|\sigma',l',u';t)=\frac{P_{\rm B}((\sigma',l',u'),(\sigma,l,u);t)}{\sum_{u''\le l''}P_{\rm B}((\sigma',l',u'),(\sigma,l'',u'');t)},
\end{eqnarray} where 
\begin{eqnarray}
P_{\rm B}((\sigma,l,u),(\sigma',l',u');t)\equiv\nonumber\\
\overline{\delta(J_{ij}-1)\delta(\sum_{k\in B_i}J_{ik}-l)
\delta(\sum_{k'\in B_j}J_{jk'}-l')P_{ij}^{\bv{J}}((\sigma,u),(\sigma',u');t)},
\end{eqnarray} with the assumption that 
the value of $P_{\rm B}((\sigma,l,u),(\sigma',l',u');t)$ does not depend on 
the chosen bonded sites $i,j$, we find that in the equilibrium case $t\ge t_{\rm eq}$,
$P_{{\rm B}}^{\sigma}(l,u|\sigma',l',u';t)=P_{{\rm B}}^{\sigma}(l,u|\sigma';t)$ holds 
because of relations (\ref{rel4}), (\ref{equality1}) and (\ref{equality2}). 
(See Appendix B for the details of the derivation of relations 
(\ref{equality1}) and (\ref{equality2}))
That is, in the equilibrium case $t\ge t_{\rm eq}$, 
approximation (\ref{rep}) becomes an identity. 

It should be noted that if $\bv{\sigma}$ at $t=0$ is 
given, $\bv{\rho}(0)$ is automatically determined. 
That is, in the case $m(0)=1$,
$\bv{\rho}(0)$ is given as
\begin{eqnarray} 
\rho_{+1,l,0}(0)&=
\left(
\begin{array}{c}
c \\ l
\end{array}
\right)
p^l(1-p)^{c-l},\\
\rho_{-1,l,0}(0)&=0, \\
\rho_{\sigma,l,u\neq 0}(0)&=0.
\end{eqnarray} 
We mention that the magnetization $m\equiv \sum_i \sigma_i/N$ 
and the energy density $e\equiv H(\bv{\sigma})/N$ 
can be expressed in terms of $\bv{\rho}$ as follows:
\begin{eqnarray}
m&=\sum_{\sigma}\sum_{u\le l}\sigma\rho_{\sigma,l,u}, \label{magd}\\
e&=-\sum_{\sigma}\sum_{u\le l}\sigma(l-2u)\rho_{\sigma,l,u}.
\end{eqnarray} 

\subsection*{A conjecture arising from the used approximation}
We consider the properties of used approximation (\ref{rep})
for the relaxation processes with the initial condition $m(0)=1$. 
Here, let $m$ and $\tau$ be the magnetization 
and the relaxation time of the model in the thermodynamic limit.
Further, let $m_{\rm d}$ and $\tau_{\rm d}$ be the magnetization 
and the relaxation time described 
by the solutions of dynamical system (\ref{dyn}).
Within the used approximation, 
we ignore the correlations of two sites between which length is far than two sites, 
although such correlations have finite relaxation times.
This leads to the conjecture that there exists a lower bound for the relaxation time $\tau$
such that $\tau_{\rm d}\le \tau$, 
which might lead to a lower bound $m_{\rm d}(t)\le m(t)$ for arbitrary $t$ with $m(0)=1$.
Partially, this conjecture is supported by the systematic work 
considering the effects of far sites in an Ising model with 
a ferromagnetic coupling \cite{Semerjian}.
In future, it will be important to provide 
mathematical proof of this conjecture.

\section{Time-dependent solutions of dynamical system (\ref{dyn})} 
We analyse the time-dependent solutions of the derived dynamical system in detail.
We consider the relaxation behaviours of the system 
from the initial condition $m(0)=1$ with fixed values of parameters $(p,T)$. 
Concretely, we numerically solve dynamical system (\ref{dyn}) 
using the forth-order Runge-Kutta method with a time step $\delta t=10^{-2}$.
In this paper, we mainly consider the system with $c=3$.
In fact, the system with $c\ge 3$ exhibits the same behaviours as the system with $c=3$ 
qualitatively, 
and the system with $c=2$ is equivalent to a one-dimensional system 
in which any phase transitions do not occur.

This section is organized as follows. 
In section 4.1, we focus on the Ising ferromagnet limit $p=1$, 
and we confirm that the results are consistent with 
those of a previous work. 
In section 4.2, we focus on the ferromagnet phase and the paramagnet phase with $p>p_c$.
In section 4.3, we focus on the Griffiths phase with $p>p_c$, 
and we find a slight discrepancy 
between the dynamical system and the MC simulations. 
In section 4.4, we focus on the Griffiths phase with $p<p_c$, 
and we find noteworthy behaviours associated with magnetization. 
It should be noted that MC simulations are carried out 
by applying the following dynamical rule. 
First, a site $i$ is randomly chosen. 
Then, the spin variable $\sigma_i$ is flipped at the probability 
$T^{\bv J}(\bv{\sigma} \to F_i\bv{\sigma})$. 
This step is repeated, and time $t=1$ is defined for such repeated $N$ steps. 
This dynamical rule is equivalent to the dynamics of the model that we consider 
in the thermodynamic limit. In all MC simulations, the system size $N$ is $10^6$, 
the number of samples for the statistical average is $10$ where 
each sample has a different realization of $\bv{J}$ and ${\rm G}(c,N)$.
\begin{figure}
\includegraphics[width=7cm,clip]{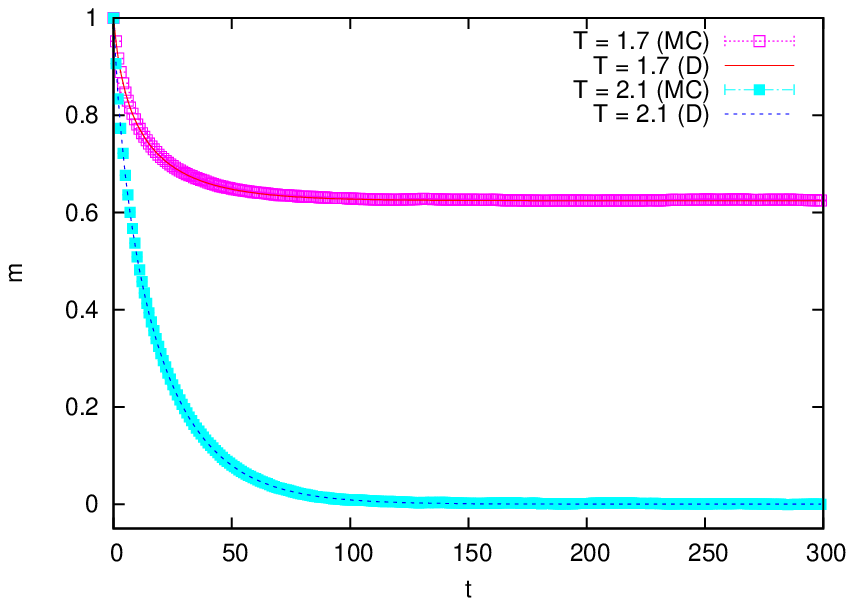}\includegraphics[width=7cm,clip]{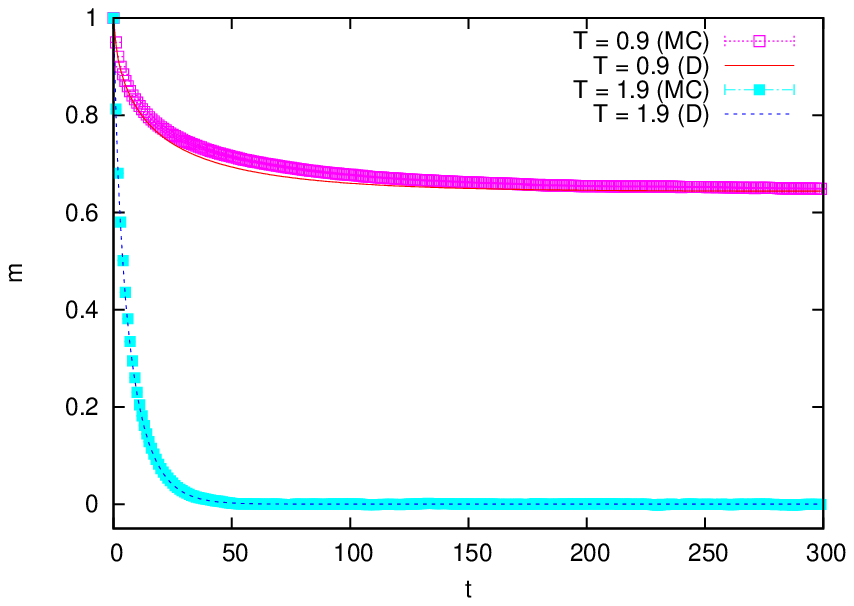}
\caption{Time-dependent magnetization described by dynamical system (\ref{dyn}) and MC simulation from the initial condition $m(0)=1$. (left) $p=1$ (right) $p=0.7$. In the figure, MC denotes MC simulations, D denotes dynamical system (\ref{dyn}).}
\label{comp1}
\end{figure}

\subsection{Ising ferromagnet limit}
Dynamical system (\ref{dyn}) with $p=1$ exactly corresponds 
to the dynamical system derived 
by the `Independent-neighbor approximation' proposed 
in the previous work \cite{Semerjian}. 
According to the work, this dynamical system can capture critical exponents 
of the dynamical critical phenomena at $T_{\rm G}(1)$. For example, the 
dynamical system can provide the relaxation time $\tau$ of the system
exhibiting $\tau\simeq\epsilon_\pm^{-\zeta}$ 
with $\zeta =1$ where $\epsilon_\pm\equiv \pm(T-T_{\rm G}(p))$. 
In the left-hand side of figure \ref{comp1}, we show two examples of the time-dependent magnetization 
described by (\ref{dyn}) together with the results of MC simulations.
It is seen that the coincidences between (\ref{dyn}) and MC simulations are quite good. 

\subsection{The ferromagnetic phase, 
and the paramagnetic phase}
Here, we focus on the behaviours with the parameter $p=0.7$.
In the right-hand side of figure \ref{comp1}, we show two examples of the time-dependent magnetization 
described by (\ref{dyn}) together with the results of MC simulations.
It is seen that the coincidences between (\ref{dyn}) and MC simulations are quite good, although there is a slight discrepancy between two results in the ferromagnetic phase $T=0.9$. In the concluding remarks, we discuss the discrepancy.

\subsection{The Griffiths phase with bond percolation clusters}
\begin{figure}
\includegraphics[width=7cm,clip]{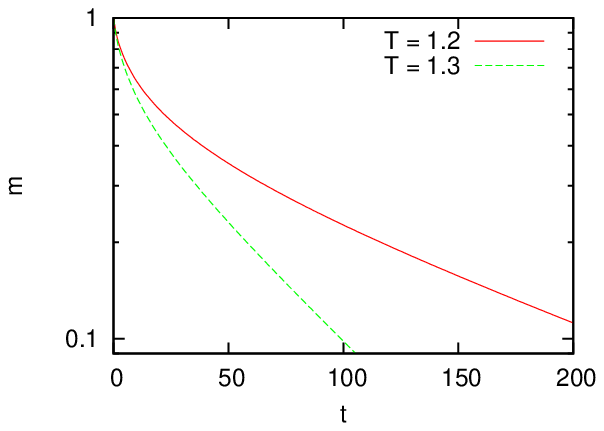}\includegraphics[width=7cm,clip]{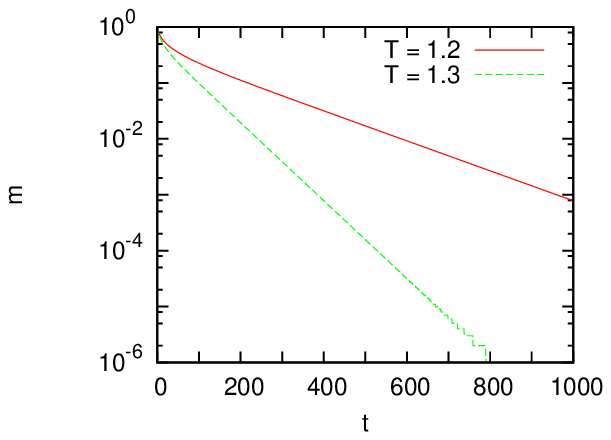}
\caption{Time-dependent magnetization described by dynamical system (\ref{dyn}) 
in the Griffiths phase with $T=1.2,1.3$ from the initial condition $m(0)=1$. $c=3$, $p=0.7$.}
\label{dgri}
\end{figure}
\begin{figure}
\centering
\includegraphics[width=7cm,clip]{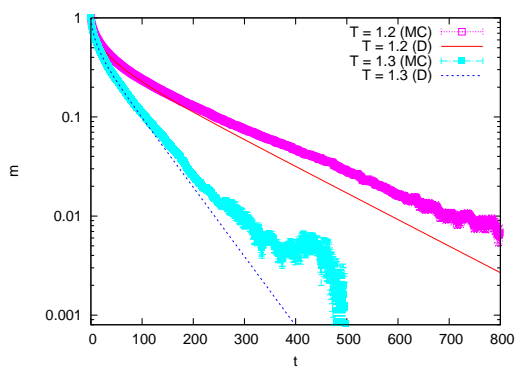}
\caption{Time-dependent magnetization described by dynamical system (\ref{dyn}) 
and MC simulations in the Griffiths phase with $T=1.2,1.3$ 
from the initial condition $m(0)=1$. $c=3$, $p=0.7$.}
\label{comp3}
\end{figure}
We focus on the behaviours with the parameter $p=0.7$.
As shown in figure \ref{dgri}, the relaxational form of the magnetization 
seems to be nonexponential as a function of time 
at an intermediate time scale; however it exhibits an exponential in the long-time limit.
In fact, the signs for nonexponential decay are found by MC simulations 
as shown in figure \ref{comp3}. 
This means that the used approximation by the analysis in this paper
fails to capture such anomalous behaviours in the long-time limit, 
although the approximation works well until an intermediate time scale.
In the concluding remarks, we discuss the validity of 
the used approximation.
It should be noted that the existence of such a slow relaxation 
has been rigorously proved in the models on $d$-dimensional lattices \cite{Maes1} 
although as far as we know, there is no rigorous proof of 
such an argument for the models on Bethe lattices.

\subsection{The Griffiths phase without bond percolation clusters}
\begin{figure}
\includegraphics[width=7cm,clip]{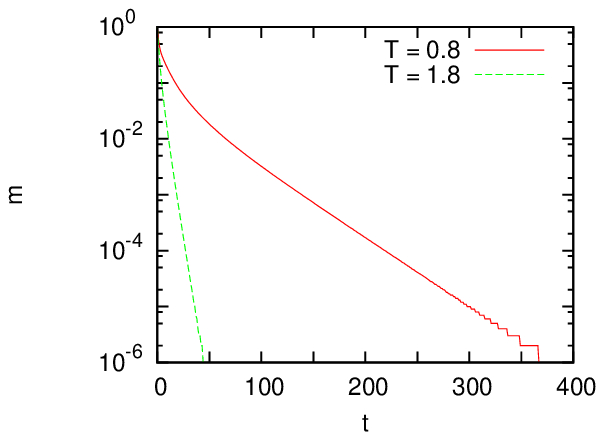}\includegraphics[width=7cm,clip]{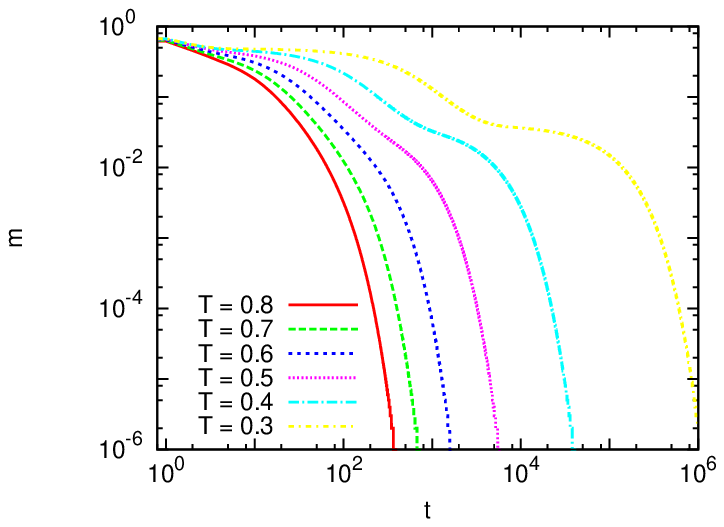}
\caption{Time-dependent magnetization described by dynamical system (\ref{dyn}) 
from the initial condition $m(0)=1$ with $c=3$, $p=1/5$.}
\label{gr2}
\end{figure}

Next, we focus on the case $p=0.2$. 
Near the transition point $T_{\rm G}(1)$, 
the relaxation form of the magnetization 
is exponential in a wide region, as seen in the left-hand side of figure \ref{gr2}. 
When the temperature is decreased to around $T=0.6$, contrary to such behaviours, 
the magnetization surprisingly shows the multi-step relaxation 
as seen in the right-hand side of figure \ref{gr2}. 
Rigorously, we use the term `$I$-step' when 
there exists the time $t_j$ for $j\in\{1,\cdots,I\}$
such that $\partial^2 (\log m)/\partial (\log t)^2|_{t=t_j}=0$.
In this sense, the three-step relaxation occurs in the case $c=3$.

\begin{figure}
\centering
\includegraphics[width=7cm,clip]{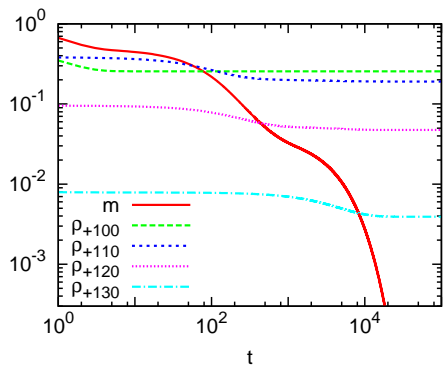}
\caption{Each quantity described by dynamical system (\ref{dyn}) 
from the initial condition $m=1$ at $T=0.4$ with $c=3$, $p=1/5$.}
\label{step}
\end{figure}
\begin{figure}
\centering
\includegraphics[width=7cm,clip]{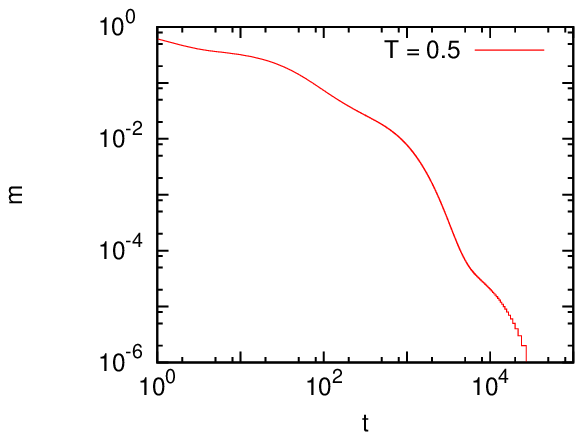}
\caption{Time-dependent magnetization described by dynamical system (\ref{dyn}) 
from the initial condition $m(0)=1$ with $c=5, p=1/10$.}
\label{c10}
\end{figure}
We expect that this observation can be related to the existence of the following terms
in (\ref{dyn});
\begin{eqnarray}
\partial_t \rho_{+1,c-i,0}(t) = -r_{+1}(c-i,0)\rho_{+1,c-i,0}(t) + A(\bv{\rho}(t)),\label{limit}
\end{eqnarray} where $i\in \{0,\cdots,c\}$, and $A(\bv{\rho}(t))$ is 
a function of $\bv{\rho}(t)$. That is, 
each variable has different relaxation time $\tau_i\simeq r_{+1}(c-i,0)^{-1}$ 
, which is almost $\exp((c-i)/T)$ with $T \ll 1$, assuming that the dependence of 
$A(\bv{\rho}(t))$ on $\rho_{+1,c-i,0}(t)$ can be ignored. 
In fact, as shown in figure \ref{step}, 
we confirm that each relaxation time $\tau_i$ is very close 
to each inflection point defined above
on the relaxation form of the magnetization.
Our conjecture is that $c+1$ number of the relaxation times
lead to $c$ number of the inflection points.
For the case $c=4,5,6$, we have confirmed that the $c$-step relaxation occurs in dynamical system (\ref{dyn}).
In the figure \ref{c10}, we show the five-step relaxation in the magnetization for the case $c=5$. 

\begin{figure}
\includegraphics[width=5cm,clip]{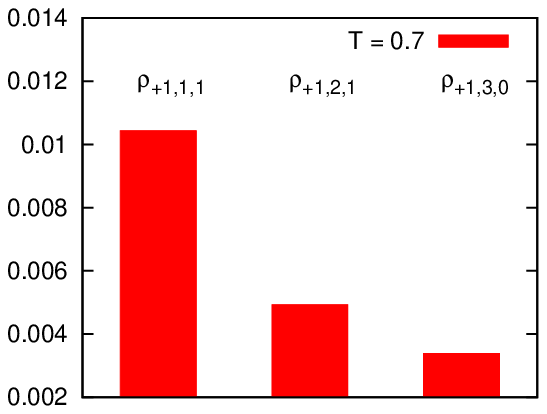}\includegraphics[width=5cm,clip]{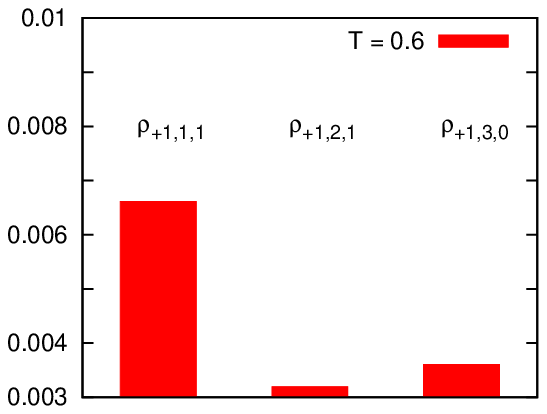}\includegraphics[width=5cm,clip]{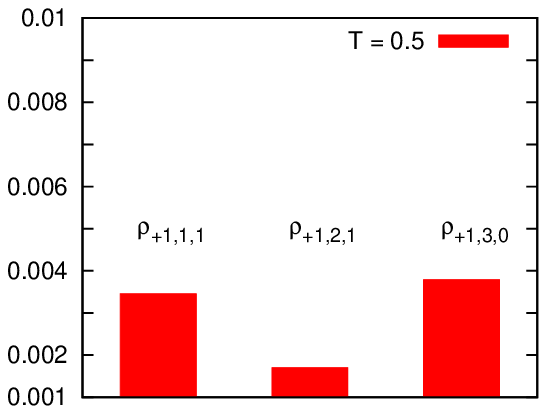}
\caption{$\rho_{+1,1,1}(t_{\rm eq})$, $\rho_{+1,2,1}(t_{\rm eq})$ and 
$\rho_{+1,3,0}(t_{\rm eq})$ depending on temperature $T$ with $c=3$, $p=1/5$.}
\label{rate}
\end{figure}

However, with only this consideration, 
we cannot understand the mechanism of the multi-step relaxations in the magnetization
because we have not yet understood what determines the threshold temperature $T_{\rm th}$ 
where a multi-step relaxation begins to occur.
We probably believe that a key fact for the appearance of the multi-step relaxations
is that the relative rate of each quantity, for example, 
$\rho_{+1,1,1}(t_{\rm eq}),\rho_{+1,2,1}(t_{\rm eq}),\rho_{+1,3,0}(t_{\rm eq})$ 
as a function of temperature $T$ is qualitatively changed 
around $T_{\rm th}$, as shown in figure \ref{rate}.
Concretely, when the temperature is decreased, $\rho_{+1,3,0}(t_{\rm eq})$ 
is increased in the manner 
$\rho_{+1,1,1}(t_{\rm eq})>\rho_{+1,2,1}(t_{\rm eq})>\rho_{+1,3,0}(t_{\rm eq})$ 
in high-temperature regions, 
$\rho_{+1,1,1}(t_{\rm eq})>\rho_{+1,3,0}(t_{\rm eq})>\rho_{+1,2,1}(t_{\rm eq})$ 
around $T_{\rm th}\simeq 0.6$, and 
$\rho_{+1,3,0}(t_{\rm eq})>\rho_{+1,1,1}(t_{\rm eq})>\rho_{+1,2,1}(t_{\rm eq})$ 
in low temperature regions. 
This leads to the conjecture 
that this qualitative change in the relative rates drives
each relaxation time $\tau_i$ become apparent in the magnetization 
as a multi-step relaxation. 
For the case $c=4,5,6$, we have confirmed that similar changes 
in the inequality to the case $c=3$ occur around 
$T_{\rm th}(p,c)$ depending on the parameters $p$ and $c$.
In spite of such observations, 
we still cannot understand how the change in the inequality is 
related to the appearance of the multi-step relaxations; 
in future, we intend to conduct a study to clarify it.
It might also be an interesting to clarify the mechanism of 
the multi-step relaxations by performing a theoretical analysis of 
dynamical system (\ref{dyn}) directly.

\begin{figure}
\centering
\includegraphics[width=7cm,clip]{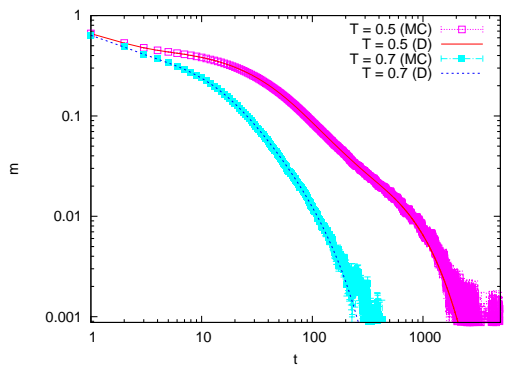}
\caption{Time-dependent magnetization described by dynamical system (\ref{dyn}) 
and MC simulation in Griffiths phase from the initial condition $m(0)=1$. $c=3$, $p=1/5$.}
\label{comp4}
\end{figure}

In figure \ref{comp4}, we show two examples of the time-dependent magnetization 
described by (\ref{dyn}) together with the results of MC simulations.
It is seen that the coincidences between (\ref{dyn}) and MC simulations 
are also quite good even if a multi-step relaxation occurs. 
This result strongly suggests that the `true' magnetization also shows 
a multi-step relaxation even in the case $c=5$ 
such as the magnetization described by dynamical system (\ref{dyn}) in figure \ref{c10}.
In this parameter, it is difficult to judge by MC simulations whether such 
behaviors occur or not.
Finally, we mention that even if such multi-step relaxations occur, 
the relaxation form of the magnetization in the long-time limit is expected to be 
exponential on the basis of the form (\ref{limit}). 
It should be noted that the existence of such an exponential form
has been rigorously proved in the models on $d$-dimensional lattices \cite{Maes2} 
although as far as we know, there is no rigorous proof of 
such an argument for the models on Bethe lattices.

\section{Concluding remarks}
In this section, we mention the relevance of our study to the other studies.
So far, as an extension of the previous work \cite{Semerjian}, 
a method called dynamical replica theory (DRT) has been already developed 
to analyse the system on Bethe lattices with quenched disorders \cite{Coolen1,Coolen2}. 
For the bond-diluted Ising model, the dynamical replica theory has already 
clarified the existence of a special temperature, 
where the relaxation of the system is qualitatively changed, and 
the qualitative change in the relative rate of some spin-field distributions \cite{Coolen1}. 
However, the concrete form of the relaxation below the special temperature 
has not been clear yet in the study \cite{Coolen1}. 
The analysis in this paper has proceeded the understanding of this phenomenon. 
That is, it has clarified that the magnetization below the special temperature 
shows multi-step relaxations. 
It is an important future study to clarify the relation 
between the method in this paper and the method of DRT
because the two method may be complementary at each other.

Before concluding this paper, we discuss the validity of the used approximation.
We mention that the used approximation
also fails to capture the anomalous relaxation dynamics 
of a kinetically constrained Ising model on a Bethe lattice, 
where heterogeneous spin configurations ordered by a kinetic constraint exist \cite{Ohta2}. 
Therefore, our conjecture is that the failure in capturing the anomalous relaxation behaviours 
below the Griffiths-paramagnetic transition temperature with bond percolation clusters 
is originated by some heterogeneous spin configurations in the relaxation dynamics, 
which might be related to rare large ordered regions.  
Furthermore, we have evidences for the conjectures that 
we can obtain better dynamical systems gradually for some systems 
by enlarging the environmental regions of a site, 
which are considered by effective variables \cite{Semerjian, Ohta2}.
Therefore, we expect that in principle, such a procedure 
can gradually extend the time span where nonexponential behaviours  
are observed as found in section 4.3, 
although such analysis is very complicated.
Finally, we mention that it is an open problem how
the used assumption related to inhomogeneous properties originated from the graph structure 
and the used approximation work at the system 
where the replica symmetry breaking occurs.

\ack 
The author wishes to thank S. Sasa for critically reading this paper, and 
providing related useful comments, discussions, and daily encouragements. 
Further, the author wishes to thank G. Semerjian 
for emphasizing the importance of the accuracy 
in equilibrium of the approximation scheme \cite{Semerjian} and providing useful comments, 
C. Maes and H. Tasaki for related useful discussions.
This work is supported by a Grant-in-Aid for JSPS Fellows.

\appendix

\section{The critical temperature $T_{\rm G}(p)$}
Let us consider a Cayley tree, which has the same structures as those of 
a random graph $\mathcal{G}\in {\rm G}(c,N)$, ignoring the loops of which length is $O(\log N)$. 
Here, we denote the generation of the Cayley tree by $g \in \{0,1,\cdots,g_{\rm max}\}$, 
where $g=0$ is assigned to the root.
For a given site in $(g-1)$-th generation, 
a bond connecting the site with a site in $g$-th generation is labeled by 
an integer $j_g\in \{1,\cdots,c-1\}$.
Thus, an arbitrary site in $g$-th generation of the Cayley tree is 
indicated by a set of integers $\bv{j}_g\equiv(j_g,\cdots,j_1,0)$.
Let $C(\bv{j}_g)$ be a set of sites defined as 
$\{(i_{g_{\rm max}},\cdots, i_{g+1},\bv{j}_g)| 1\le i_k\le c-1,g+1\le k \le g_{\rm max}\}$ 
where $(i,k,\bv{j}_g)\equiv (i,k,j_g,\cdots,j_1,0)$.
The sites in $C(\bv{j}_g)$ can be regarded as sites on a Cayley tree ${\rm T}_{\bv{j}_g}^{\bv{j}_{g-1}}$ 
with the root being site $\bv{j}_g$.
Then, for a realization of diluted-bonds $\bv{J}$, 
let us consider a partition function $Z_{\bv{j}_g}^{\bv{J}}(\sigma)\equiv 
\sum_{i\in C(\bv{j}_g)}\sum_{\sigma_i}\delta(\sigma_{\bv{j}_g}-\sigma)
\exp(-\beta H^{\bv{J}}((\sigma_i)_{i\in C(\bv{j}_g)}))$ where $\beta\equiv 1/T$.
Here, we define $\exp(\beta 2h_{\bv{j}_g}^{\bv{J}})\equiv Z_{\bv{j}_g}^{\bv{J}}(-1)/Z_{\bv{j}_g}^{\bv{J}}(1)$.
Then, we can obtain the recursive equation in terms of the distribution 
$Q_{\bv{j}_{g}}(h)\equiv \overline{\delta (h-h_{\bv{j}_g}^{\bv{J}})}$ as follows:
\begin{eqnarray}
Q_{\bv{j}_{g}}(h) = H([Q_{\bv{j}_{g+1}}(h)]),\label{rec}\\
H([Q(h)])=\int dh_i Q(h_i)\overline{\delta 
(h-\sum_{i=1}^{c-1}u(J_{(i,\bv{j}_g),\bv{j}_g},h_i)},
\end{eqnarray} 
where $u(J,h)=\beta^{-1} \tanh^{-1}[\tanh(\beta J)\tanh(\beta h)]$.
 
It should be noted that by solving recursive equation (\ref{rec}) 
with almost all values of initial conditions $Q_{\bv{j}_g}(h)$ with $g=g_{\rm max}$, 
$Q_{\bv{j}_g}(h)$ for arbitrary $\bv{j}_g$ with $g \ll g_{\rm max}$ 
becomes a solution $Q(h)$ satisfying the following equation
\begin{eqnarray}
Q(h)=H([Q(h)]).\label{fix}
\end{eqnarray}
According to the previous study \cite{Laumann}, 
for a sufficiently high temperature, the solution $Q(h)$ is $\delta(h)$.
When the temperature $T$ is decreased from a sufficiently high temperature 
for a fixed $p$, we can obtain the critical temperature $T_{\rm G}(p)$ 
where the point $h_{\bv{j}_{g}}(\equiv \int dh Q_{\bv{j}_{g}}(h)h)=0$ loses 
linear stability
in the recursive equation between $h_{\bv{j}_{g}}$ and $h_{\bv{j}_{g+1}}$, 
which is derived from recursive equation (\ref{rec}). 
From this, we obtain the relation
$1/T_{\rm G}(p)=\tanh^{-1}(p_{\rm c}/p)$, where $p_{\rm c}=1/(c-1)$ is 
a threshold value in a bond-percolation problem. 
That is, there are bond percolation clusters in the system with $p>p_{\rm c}$, 
no bond percolation clusters in the system with $p<p_{\rm c}$.
Concretely, $T_{\rm G}(1)=1.820478$ for $c=3$, and for $c<3$, 
any phase transitions do not occur in the model.
 
\section{Key relations in equilibrium}
Let us consider site $i$ and sites $j\in B_i$ in a random graph $\mathcal{G}\in {\rm G}(c,N)$. 
Here, we assume that the sites $j$, respectively, can be regarded as the root 
of the Cayley tree ${\rm T}_j^i$, of which generation is not increased to the direction of site $i$.

Under this condition, when we ignore site $i$ for site $j\in B_i$, we can regard 
$Z_{j}^{\bv{J}}(\sigma)/\sum_{\sigma'}Z_{j}^{\bv{J}}(\sigma')$
as the probability that $\sigma_j$ takes a value $\sigma$.
Here, let us restate the bonded sites of site $i$, $\{j_1,\cdots,j_{l_i}\}$, 
and the unbonded sites of site $i$, $\{j_{l_i+1},\cdots,j_{c}\}$.
Using this representation, we can obtain the concrete expression of the probability 
$P_i^{\bv{J},\rm{eq}}(\sigma,u)$
that with a realization of $\bv{J}$, 
$(\sigma_i,u_i)$ takes $(\sigma,u)$ in equilibrium as follows:
\begin{eqnarray}
P_i^{\bv{J},\rm{eq}}(\sigma,u)=\frac{1}{N_0^{\bv{J}}}\exp(-\beta\sigma (l_i-2u))\nonumber\\
\sum_{\sum_{b\le l_i}\sigma_{j_b} = l_i-2u}
\prod_{b = 0 }^{l_i}\frac{Z_{j_b}^{\bv{J}}(\sigma_{j_b})}{\sum_{\sigma'}Z_{j_b}^{\bv{J}}(\sigma')}
\prod_{b'= l_i+1}^c\frac{\sum_{\sigma'}Z_{j_{b'}}^{\bv{J}}(\sigma')}{\sum_{\sigma'}Z_{j_{b'}}^{\bv{J}}
(\sigma')},
\end{eqnarray} where $N_0^{\bv{J}}$ is determined by $\sum_{\sigma,u}P_i^{\bv{J},{\rm eq}}(\sigma,u)=1$.

Let us consider the probability $\rho_{\sigma,l,u}^{\rm{eq}}$ defined as
\begin{eqnarray}
\rho_{\sigma,l,u}^{\rm{eq}}\equiv \overline{\delta(\sum_{j\in B_i}J_{ij}-l)P_i^{\bv{J},\rm{eq}}(\sigma,u)},
\end{eqnarray} which corresponds to $\rho_{\sigma,l,u}(t\ge t_{\rm eq})$.
Here, we have assumed that 
$\rho_{\sigma,l,u}^{\rm eq}$ does not depend on the chosen site $i$, and 
$\overline{Z_{j}^{\bv{J}}(-1)/Z_{j}^{\bv{J}}(1)}$ does not depend on site $j$. 
When we define $\exp(\beta 2h_{\rm av})\equiv \overline{Z_{j}^{\bv{J}}(-1)/Z_{j}^{\bv{J}}(1)}$, 
using $Q(h)$ which has been already obtained in (\ref{fix}), 
we can find $\exp(\beta 2h_{\rm av})=\int dh Q(h)\exp(\beta 2h)$.
Then, we obtain the following exact expression of the probability $\rho_{\sigma,l,u}^{\rm eq}$:
\begin{eqnarray}
\rho_{\sigma,l,u}^{\rm{eq}}=\frac{1}{N_v}
\left(
\begin{array}{c}
c \\ l
\end{array}
\right)p^l(1-p)^{c-l}
\left(
\begin{array}{c}
l \\ u
\end{array}
\right)\nonumber\\
\exp( -\beta\sigma (l-2u))\exp(u\beta 2h_{\rm av})(1 + \exp(\beta 2h_{\rm av}))^{c-l},\label{key1}
\end{eqnarray} where we used relation (\ref{rel}) and determined $N_v$ by the condition
$\sum_{\sigma}\sum_{u\le l}\rho_{\sigma,l,u}^{\rm eq}=1$.

Next, we discuss the joint probability $P_{ij}^{{\bv{J}},{\rm eq}}((\sigma,u),(\sigma',u'))$ 
that in equilibrium, with a realization of $\bv{J}$, $(\sigma_i,u_i)$ and $(\sigma_j,u_j)$ 
take $(\sigma,u)$ and $(\sigma',u')$, respectively. 
Let us consider the probability 
\begin{eqnarray}
P_{\rm B}^{\rm{eq}}((\sigma,l,u),(\sigma',l',u'))
\equiv \nonumber\\
\overline{\delta(J_{ij}-1)\delta(\sum_{m\in B_i}J_{im}-l)\delta(\sum_{m\in B_j}J_{jm}-l')
P_{ij}^{{\bv{J}},{\rm eq}}((\sigma,u),(\sigma',u'))},
\end{eqnarray} which corresponds to $P_{\rm B}((\sigma,l,u),(\sigma',l',u');t\ge t_{\rm eq})$ 
assuming  $P_{\rm B}^{\rm{eq}}((\sigma,l,u),(\sigma',l',u'))$ does not depend chosen bonded sites $i,j$.
In the same way as $\rho_{\sigma,l,u}^{\rm eq}$, we obtain the exact expression as follows:
\begin{eqnarray}
P_{\rm B}^{\rm{eq}}((\sigma',l',u'),(\sigma,l,u))=\frac{1}{N_{e}}
\left(
\begin{array}{c}
c-1 \\ l'-1
\end{array}
\right)
\left(
\begin{array}{c}
c-1 \\ l-1
\end{array}
\right)
p^{l'-1}(1-p)^{c-1-(l'-1)}\nonumber\\p^{l-1}(1-p)^{c-1-(l-1)}
\left(
\begin{array}{c}
l'-1 \\ u'-\delta(\sigma+1)
\end{array}
\right)
\left(
\begin{array}{c}
l-1 \\ u-\delta(\sigma'+1)
\end{array}
\right) p\exp(-\beta\sigma\sigma')\nonumber \\
\exp(-\beta\sigma' (l'-1-2(u'-\delta(\sigma+1)))
-\beta\sigma(l-1-2(u-\delta(\sigma'+1)))\nonumber\\
+ (u'- \delta(\sigma+1) + u - \delta(\sigma'+1))\beta 2h_{\rm av})(1 + \exp(\beta 2h_{\rm av}))^{2c-l-l'},\label{key2}
\end{eqnarray} where $N_e$ is determined by $\sum_{\sigma'}\sum_{\sigma}\sum_{u\le l}\sum_{u'\le l'}
P_{\rm B}^{\rm eq}((\sigma',l',u'),(\sigma,l,u))=1$.
It should be noted that from expressions (\ref{key1}) and (\ref{key2}), 
we derive key relations
\begin{eqnarray}
\frac{P_{\rm B}^{\rm{eq}}((+1,l',u'),(\sigma,l,u))}{\sum_{l'',u''}P_{\rm B}^{\rm{eq}}((+1,l',u'),(\sigma,l'',u''))}
=\frac{(l-u)\rho_{\sigma,l,u}^{\rm{eq}}}{\sum_{l'',u''} (l''-u'')\rho_{\sigma,l'',u''}^{\rm{eq}}},\label{equality1}\\
\frac{P_{\rm B}^{\rm{eq}}((-1,l',u'),(\sigma,l,u))}{\sum_{l'',u''}P_{\rm B}^{\rm{eq}}((-1,l',u'),(\sigma,l'',u''))}
=\frac{u\rho_{\sigma,l,u}^{\rm{eq}}}{\sum_{l'',u''} u''\rho_{\sigma,l'',u''}^{\rm{eq}}}.\label{equality2}
\end{eqnarray}

\end{document}